\documentstyle[amstex,amssymb,preprint,aps]{revtex}

\begin{document}
\title{Nonlocality tests of Bell Inequalities and of Hardy's ''ladder theorem''
without ''supplementary assumptions''}
\author{M. Barbieri, F. De Martini, G. Di Nepi and P. Mataloni}
\address{Dipartimento di Fisica and \\
Istituto Nazionale per la Fisica della Materia\\
Universit\`{a} di Roma ''La Sapienza'', Roma, 00185 - Italy}
\maketitle

\begin{abstract}
We have experimentally tested the non local properties of the states
generated by a high brilliance source of entanglement which virtually allows
the direct measurement of the full set of photon pairs created by the basic
QED process implied by the parametric quantum scattering. Standard Bell
measurements and Bell's inequality violation test have been realized over
the entire cone of emission of the degenerate pairs. By the same source we
have verified the Hardy's ladder theory up to the 20th step and the
contradiction between the standard quantum theory and the local realism has
been tested for $41\%$\ of entangled pairs.

\end{abstract}


Entanglement, ''{\it the characteristic trait of quantum mechanics}''{\it \ }%
according to Erwin Schroedinger, is playing an increasing role in nowadays
physics \cite{1}. Since the EPR discovery in 1935 followed by a many decades
long endeavour ending with the emergence of\ the Bell's inequalities and
with the experiment by Alain Aspect, entanglement is considered as the
irrevocable signature of quantum nonlocality, i.e. the scientific paradigm
recognized as the fundamental cornerstone of our yet uncertain understanding
of the Universe \cite{2,3,4}. In recent years the violation of these
inequalities has been succesfully tested many times by optical experiments,
mostly involving polarization entangled photons generated by Spontaneous
Parametric Down Conversion (SPDC) in a nonlinear (NL)\ crystal. In addition,
an other nonlocality test not involving inequalities was proposed years ago
by Lucien Hardy's \cite{5} and soon realized experimentally by a SPDC
process \cite{6}.

The present work reports yet another nonlocality test both of the standard
Bell configuration and of the Hardy's no-inequality scheme. The novelty of
this experiment consists of the peculiar spatial properties of the output $%
{\bf k}-$ vector distribution generated by the SPDC source implied by the
present scheme \cite{7,8}. As shown in the paper, this source allows, at
least in principle, the coupling to the output detectors of the full set of
optical modes carrying the particle pairs involved in the EPR measurement.
In other words, all entangled pairs created over the entire set of
wavevectors allowed by phase matching can virtually be detected. Since then
the detected emission process is entirely ''quantum'', i.e. not affected by
any previous ''classical'' manipulation, such as wavelength $(\lambda )$ of
wavevector $(k)$ filtering, e.g. by flters and/or limiting pinholes, the new
scheme allows in principle the realization of the necessary premises
underlying the original formulation of the ''EPR Paradox'' \cite{9}. Indeed
all nonlocality tests performed so far were affected by a quantum-efficiency 
$(QE)$ ''loophole'' expressing the overall lack of detection of\ all couples
of entangled photons generated by the EPR source \cite{10}. This effect is
ascribable either to the limited $QE$ of the detectors (''{\it detection} 
{\it QE}'': $dQE$) \cite{11} and to the loss of the pairs that, created by
the underlying QED\ quantum process, could not reach the detectors for
geometrical reasons (''{\it collection QE}'': $cQE$). Note that, while $dQE$
can be of the order of $10^{-1}$ for normal detectors in the visible range,
the $cQE$ contribution has been always typically less than $10^{-5}$. As
shown later, this filtering truncation of the distribution of the emitted
entangled pairs necessarily results in a mixed character of the detected
state, at variance with the original EPR assumptions \cite{9}. In principle
our scheme relieves the need for the {\it fair sampling} and {\it no
enhancement} ''supplementary assumptions'' in the analysis of the test
outcomes, a condition long advocated by John Bell himself and never realized
in practice \cite{3,10,11,12,12bis}.

A detailed description of the high brilliance source of entanglement was
already given in previous papers \cite{7,8}. Pairs of horizontally ($H$)
polarized SPDC photons are emitted at wavelength $\lambda $ over the surface
of the phase matching cone of a thin ($0.5mm$) type I BBO crystal which is
excited by a cw vertically ($V$) polarized $Ar^{+}$ laser beam ($\lambda
_{p}=\lambda /2$) (see Fig. 1). A spherical mirror $M$ with radius $R$,
placed at a distance $d=R$ \ from the crystal, reflects back both the pump
and the photons. By a zero-order $\lambda /4$ waveplate (wp) placed between $%
M\ $and the BBO the $H\longrightarrow V$ transformation for the $\lambda $
photons polarization is performed while the pump beam is left in its
original polarization state. This excites an identical SPDC process over a
new radiation cone which is spatially and temporally indistinguishable from
the previous one. The state of the overall radiation is then expressed by
the entangled state: 
\begin{equation}
|\Phi \rangle =\frac{1}{\sqrt{2}}\left( |HH\rangle +e^{i\phi }|VV\rangle
\right)  \label{phi}
\end{equation}
with phase $\phi $\ ($0\leq \phi \leq \pi $)\ reliably controlled by
micrometric displacements of $M$. By a positive lens the overall conical{\it %
\ }emission distribution is transformed into a cylindrical{\it \ }one whose
transverse circular section, spatially selected by an annular mask,
identifies the Entanglement-ring{\it \ }(E-ring) (Fig. 1). After division of
the ring along a vertical axis, the two resulting equal portions are
detected at sites ${\cal A}$ and ${\cal B}$ within a bandwidth $\Delta
\lambda =6nm$. More than $4\times 10^{3}\sec ^{-1}$ coincidences are
measured at a pump power $P\simeq 100mW$ over the entire E-ring.

We may analyze the structural characteristics of the quantum state of any
photon pair generated by our source by accounting first for the excited
electromagnetic modes which, in our case are grouped in correlated pairs by
the $3-$wave SPDC interaction. Assume that each SPDC ${\bf k}${\bf -}cone is
represented by a linear superposition of correlated pairs of e.m. modes $(%
{\bf k}_{1},{\bf k}_{2})$. Since only one pair of photons is detected at the
output of the source, each mode corresponds to a Fock $2-$mode product-state
that can be either $|0,0\rangle $ or $|1_{H},1_{H}\rangle $ or $%
|1_{V},1_{V}\rangle $. Accordingly, we can express the overall
entangled-state by the quantum superposition: 
\begin{multline}
|\Phi \rangle =\int d{\bf k}_{1}d{\bf k}_{2}\left( |1_{H},1_{H}\rangle _{%
{\bf k}_{1}{\bf k}_{2}}|0_{V},0_{V}\rangle _{{\bf k}_{1}{\bf k}%
_{2}}+e^{i\phi }|0_{H},0_{H}\rangle _{{\bf k}_{1}{\bf k}_{2}}|1_{V},1_{V}%
\rangle _{{\bf k}_{1}{\bf k}_{2}}\right)  \label{sum} \\
\bigotimes_{({\bf k}_{1}^{\prime }{\bf k}_{2}^{\prime })\neq ({\bf k}_{1}%
{\bf k}_{2})}|0_{H},0_{H}\rangle _{{\bf k}_{1}^{\prime }{\bf k}_{2}^{\prime
}}|0_{V},0_{V}\rangle _{{\bf k}_{1}^{\prime }{\bf k}_{2}^{\prime }}
\end{multline}
This state should indeed express the exact form of the single photon-pair
output state of the source if the{\it \ }full set of mode pairs could be
coupled to the detectors ${\cal A}$ and ${\cal B}$. Indeed, it is not
difficult to conceive an ideal experiment (an approximate one is in fact in
progress in our laboratory) by which the full set of modes at any
wavelength, either degenerate or non-degenerate can be coupled to the
detectors without any geometrical or frequency constraint, i.e. without any
spatial or $\lambda $-filtering. In practice, in this case limitations for
an overall full particle detection should come from the limited $\lambda -$%
extension of the photocathode $dQE^{\prime }s$\ and of the performance of
the optical components (mirrors, lenses etc.). Nevertheless, this should not
affect {\it in principle} the structural character of the output
entangled-state. As said this condition gets rid of the {\it fair sampling}
and {\it no enhancement} ''supplementary assumptions'' in the analysis of
the test outcomes \cite{3,10}.

Note that in any typical SPDC-based experiment the set of mode pairs coupled
to the detectors are drastically reduced by the use of very narrow
spatial-filtering pinholes\ \cite{13,14} in order to realize the
photodetection over a single pair of correlated ${\bf k}-$vectors belonging
to the distribution appearing in Eq.(\ref{sum}). However this operation
cannot be realized but within a mode uncertainty $\Delta {\bf k}$ because of
the inescapable effect of diffraction{\bf .} In these conditions there is a
definite probability that only one photon in a pair passes through the
spatial filter while the other one is intercepted. A similar effect can be
ascribed to any frequency filtering operation as well. As a consequence,
this drastical truncation implies necessarily a {\it mixed character} of the
output entangled state\cite{14bis}. These considerations lead to the
quasi-purity of the generated output state. The state purity condition may
be simply analyzed as follows. The well known unitary character of the SPDC\
quantum operator $\hat{S}$\ \ assures that the purity of the input state
implies also the purity of the output state:{\it \ }$|\Phi \rangle _{out}=%
\hat{S}|\Phi \rangle _{in}$ \cite{15}. Adopting the common hypotesis of a
undepleted ''classical'' pump beam, the input \ pure state is expressed by
the overall vacuum-state character of the full set of input modes acted upon
by the SPDC\ process: $|\Phi \rangle _{in}\equiv |vac\rangle $. Within the
single-pair emission approximation, the output pure\ state is found: $|\Phi
\rangle _{out}$ $\simeq |\Phi \rangle +|vac\rangle $, viz. consists of the
sum of the state given by Eq.(\ref{sum}) and of \ the vacuum-state
expressing the non realization of the QED scattering process. As a
consequence, $|\Phi \rangle $ given by Eq.(\ref{sum}) is not{\it , }strictu
sensu{\it ,} a pure state but one out of a two components mixture.\ However,
in the common case of a conditional experiment{\it \ }where the overall
registration system is activated by a trigger pulse elicited by the source
itself, the output state $|\Phi \rangle $ may be considered a
''post-selected'' pure state. This last condition is often referred to as
expressing the ''conditional purity'' of the output state.

The experimental interference pattern, with coincidence visibility $V\geq
94\%$, shown in Fig. 2a gives a strong indication of the entangled nature of
the Bell state $|\Phi ^{-}\rangle $, ($\phi =\pi $) over the entire emission
cone at $\lambda =2\lambda _{p}$ In this condition it is possible to
evaluate that $cQE$ is enhanced of a factor $\geqq 70$ with respect the
standard pinhole configuration. The dotted line corresponds to the limit
boundary between the quantum and the classical regimes \cite{16} while the
theoretical continuous curve expresses the ideal interferometric pattern
with maximum visibility{\it :} $V=1$. By performing the standard
Bell-inequality test we have evaluated the non locality parameter $S$ \cite
{3}. The measured value $S=2.5564\pm .0026$ \cite{7}, obtained by
integrating the data over $180s$, corresponds to a violation as large as $%
213 $ standard deviations respect to the limit value $S=2\ $implied by local
realistic{\it \ }theories.

Hardy's theorem represents an alternative proof of nonlocality \cite{5,6}.
It is obtained in the case of non maximally entangled states of two spin $%
1/2 $ particles: 
\begin{equation}
|\Phi \rangle =\alpha |H,H\rangle -\beta |V,V\rangle \text{, \ \ }(0\leq
\alpha \leq \beta ;\ \alpha ^{2}+\beta ^{2}=1)\text{.}  \label{nonmax}
\end{equation}
We have realized these states by inserting a zero-order $\lambda _{p}/4$ wp
between $M\ $and the BBO, intercepting only the UV\ beam (Fig. 1). In our
system, by rotating the UV\ wp by an angle $\theta _{p}$, the back-reflected
UV\ pump beam experiences a polarization rotation of $2\theta _{p}$ respect
to the optical axes of the NL\ crystal slab. As a consequence, the emission
efficiency of the $|H,H\rangle $ cone is decreased by a coefficient $\propto
\cos ^{2}2\theta _{p}$. By adjusting $\theta _{p}$ in the range $0-\pi /4$,
the degree of entanglement $\gamma =\alpha /\beta $ can be continuously
tuned between $0$ and $1$.

A full presentation of Hardy's theorem can be found in Ref \cite{6}. For two
photons in the state (\ref{nonmax}), the polarization measurements performed
along $K+1$ possible directions at sites ${\cal A}$ and ${\cal B}$ of Fig. 1
give a corresponding set of propositions ${\bf A}_{k}$, $k=0,..,K$ which
imply ${\bf A}_{0}\Longrightarrow {\bf A}_{1}\Longrightarrow
...\Longrightarrow {\bf A}_{K}$, with the further condition ${\bf A}%
_{0}\nRightarrow {\bf A}_{K}$. It can be demonstrated that the fraction of
pairs $P_{K}$ with non local properties increases with $K$ and is a function
of the entanglement degree $\gamma .$ For each value of $K$,\ a proper $%
\gamma $.exists which maximizes $P_{K}.$ Hardy's ladder proof is purely
logical and doesn't involve inequalities. However inequalities are necessary
as a quantitative test in a real experiment in order to avoid the conceptual
problems associated to the realization of a {\it nullum experiment }\cite{6}%
. A proper inequality, violated by quantum theory, can be derived by
combining Hardy's theorem with Clauser-Horne inequality \cite{2,3}. It
consists of the measurement of $2K+2$ joint detection probabilities $%
P(\theta _{A},\theta _{B}),$ where $\theta _{A}$, $\theta _{B}$ are the
angular settings of polarizers on sites ${\cal A}$ and ${\cal B}$ in Fig.1: 
\begin{equation}
P(\theta _{K},\theta _{K})\leq P(\theta _{0},\theta _{0})+\sum_{k=1}^{K} 
\left[ P(\theta _{k},\theta _{k-1}^{\bot })+P(\theta _{k-1}^{\bot },\theta
_{k})\right] ={\cal P}  \label{ineq}
\end{equation}
where $\theta _{k}=(-1)^{k}\arctan (\gamma ^{k+\frac{1}{2}}),$ $\theta
_{k}^{\bot }=\theta _{k}+\frac{\pi }{2},$with $k=0,...,K$, and $P(\theta
_{K},\theta _{K})=P_{K}$.

The experimental observation of the inequality violation becomes more and
more difficult as $K$ increases because of an eventually unperfect
definition of the state and of the experimental uncertainties associated to
all the $2K+2$ measurements. The experiments realized so far were performed
only for low values of $K$, in particular for $K\leq 3$ \cite{6,17,18}. The
above described source possesses unique characteristics for this experiment.
In fact, it allows the direct generation of non maximally entangled states
without postselection. Moreover, the particular configuration of ''single
arm'' interferometer guarantees a very high phase stability for long periods
($>1hr$). Finally, the high brilliance character of the source allows to
accumulate large sets of statistical data in a short measurement time $%
\Delta T$ also with a relatively low UV\ pump power. By taking advantage
from all these properties of our source, we could successfully test Hardy's
ladder proof for large values of $K.$

The experiment, realized for $K=4$, $5$, $10$, $20$, has given the following
violations of the inequality (\ref{ineq}):

$K=4$ $(\Delta T=60\sec )$: $P_{4}=0.2586\pm 0.0041$; ${\cal P}=0.1213\pm
0.0022$. Inequality violated for $30\sigma $.

$K=5$ $(\Delta T=60\sec )$: $P_{5}=0.3152\pm 0.0050$; ${\cal P}=0.1184\pm
0.0022$. Inequality violated for $37\sigma $.

$K=10$ $(\Delta T=120\sec )$: $P_{10}=0.3402\pm 0.0045$; ${\cal P}=0.2288\pm
0.0015$. Inequality violated for $26\sigma $.

$K=20$ $(\Delta T=180\sec )$: $P_{20}=0.4132\pm 0.0053$; ${\cal P}=0.2439\pm
0.0016$. Inequality violated for $21\sigma $.

The probabilities of each outcome for all the $42$ polarization settings of $%
K=20$ are reported in Table 1. These have been obtained by normalizing the
coincidence measurements to the sum of coincidence rates measured in the
basis $\left| HH\right\rangle $ and $\left| VV\right\rangle $.

The count rates for each value of $P_{K}$ are plotted in Fig. 2b as a
function of $K$. We report for comparison the results obtained in the
experiment of ref \cite{6}. The thoretical curve shown in the same Figure
indicates a very slow convergence to the asymptotic value $P_{K}=0.5$.

Finally, additional measurements of $P_{K}$ are plotted as a function of $%
\gamma $ for $K=4$, $5$, $10$, $20$ in Fig. 3. The angle $\theta _{K}$\ has
been calculated for each value of $\gamma $ by using .the above given
expression. The agreement with the theory appears very good.

In summary, we have presented two different experimental tests of quantum
nonlocality realized by a high brilliance source of polarization
entanglement. The value of the {\it collection Quantum Efficiency} $(cQE)$
realized by the present system is about $2$\ order of magnitude larger than
for all previous experiments. We have obtained in these conditions a $%
213\sigma $ Bell inequality violation. Furthermore, in virtue of the very
large overall efficiency of the source, within the framework of the Hardy's
ladder theory, a contradiction between standard quantum theory and local
realism has been attained by for a fraction as large as $41\%$\ of the
entangled photon pairs and as many as $20$ steps of the ladder have been
realized.

This work was supported by the FET European Network on Quantum Information
and Communication (Contract IST-2000-29681: ATESIT), MIUR
2002-Cofinanziamento and PRA-INFM\ 2002 (CLON).

\centerline{\bf Figure Captions}

\vskip 8mm

\parindent=0pt

\parskip=3mm

Fig. 1- Layout of the high brilliance{\it \ }source of polarization
entanglement. The dimension of the annular mask are $D=1.5cm$, $\delta
=.07cm $.

Fig. 2- (a) Measurement of the polarization entanglement for the state $%
\left| \Phi ^{-}\right\rangle =\frac{1}{\sqrt{2}}\left( \left|
H,H\right\rangle -\left| V,V\right\rangle \right) $ obtained by varying the
angle $\theta _{A}$ on site ${\cal A}$ in the range ($45%
{{}^\circ}%
-135%
{{}^\circ}%
$), having kept fixed the angle $\theta _{B}=45%
{{}^\circ}%
$ on site ${\cal B}.$ (b) Plot of $P_{K}$ against $K$. Black circles:
experimental results for $K=4$, $5$, $10$, $20$ (error bars are lower than
the dimension of the corresponding experimental points). White circles:
experimental results obtained in ref. \cite{6}.

Fig. 3- Plots of $P_{4}$, $P_{5}$, $P_{10}$, $P_{20}$ as a function of $%
\gamma $. The solid curves represent the theoretical predictions. The error
bars are lower than the dimension of the corresponding experimental points.

Tab. 1- Experimental joint probabilities for $K=20$.


\begin{references}
\bibitem{1}  E. Schroedinger, {\it Proc. Cambridge Phil. Soc}. {\bf 31}, 555
(1935).

\bibitem{2}  J. S. Bell {\it Physics}.{\bf 1}, 165 (1964);

\bibitem{3}  J. F. Clauser, M. A. Horne, A. Shimony and R. A. Holt, {\it %
Phys. Rev. Lett. }23, 880 (1969); J. F. Clauser and M. A. Horne, Phys. Rev.
D 10, 526 (1974).

\bibitem{4}  A. Aspect, P. Grangier, G. Roger, {\it Phys. Rev. Lett.} {\bf 47%
}, 460 (1981); A. Aspect, P. Grangier, G. Roger, {\it Phys. Rev. Lett }{\bf %
49}, 91 (1982); A. Aspect, J. Dalibard, G. Roger, {\bf 49}, 1804 (1982). The
latter paper reports on the very first effect of aleatory scrambling
performed on the linear polarization correlation of photon pairs by time
vaying analyzers.

\bibitem{5}  L. Hardy, {\it Phys. Rev. Lett.} {\bf 71}, 1665 (1993).

\bibitem{6}  D. Boschi, S. Branca, F. De Martini, L. Hardy, {\it Phys. Rev.
Lett. }{\bf 79}, 2755 (1997).

\bibitem{7}  G. Giorgi, G. Di Nepi, P. Mataloni and F. De Martini, {\it %
Laser Physics}, {\bf 13}, 350 (2003); M. Barbieri, F. De Martini, G. Di
Nepi, P. Mataloni, {\it Phys. Rev. Lett.} 92, 177901 (2004).

\bibitem{8}  M. Barbieri, F.De Martini, G.Di Nepi, P.Mataloni, G.M.D'Ariano,
C.Macchiavello, {\it Phys. Rev. Lett.}, {\bf 91}, 227901 (2003).

\bibitem{9}  A.Einstein, B.Podolsky and N.Rosen, Phys. Rev. {\bf 47}, 777
(1935) and in: {\it Physik und Realit\"{a}t, }Journ.of Franklin Inst{\it . 
{\bf 221}, 313 (1936). }According to Einstein{\it \ }the concept of quantum 
{\it wavefunction}, is justified only in a {\it statistical} sense and is
not applicable to {\it single systems}. Otherwise it would lead to a
paradox. Note that a {\it pure} entangled state is the necessary premise of
the original formulation of the EPR\ paradox.

\bibitem{10}  F. Selleri, {\it Weak and strong Bell type inequalities,} in 
{\it Fisica Cu\`{a}ntica y Realidad}, ed. by C. Mataixe and A. Rivadulla
(Ed. Complutense, Madrid, 2002).

\bibitem{11}  E. Santos, {\it Phys. Rev. Lett.} 66,1388 (1991) N.D. Mermin,
in {\it New Techniques and Ideas in Quantum Measurement} (The N.Y.Academy of
Sciences, New York, 1986);\ A. Garuccio, {\it Phys. Rev. A}, {\bf 52}, 2535
(1995); E. S. Fry, T. Walther and S. Li, {\it Phys. Rev. A}, {\bf 52}, 4381
(1995).

\bibitem{12}  The condition $cQE=1$ was advocated by J. Bell (A. Aspect,
private comm. to FDM; J. Bell, {\it Atomic-cascade photons and quantum
mechanical nonlocality}, {\it Comments on Atomic and Molecular Physics}, 
{\bf 9}, 121(1980) reprinted in {\it Speakable and Unspeakable in Quantum
Mechanics} (Cambridge U. Press, 1987)). There the Aspect's experiment, in
which no momentum correlation between entangled particles exists, was
considered. In the case of SPDC tests, in which the spatial correlation is
imposed by phase matching condition, the diffraction effects are determined
by the finite thickness of the NL crystal and by the finite pump beam
transverse profile.

\bibitem{12bis}  Here we deal with the supplementary assumptions which are
necessary because of the limited value of $cQE$.

\bibitem{13}  D. Klyshko, {\it Photons and Nonlinear Optics}, (Gordon and
Breach, New York, 1988).

\bibitem{14}  P. G. Kwiat, K. Mattle, H. Weinfurter, A. Zeilinger, A. V.
Sergienko and Y. Shih, {\it Phys. Rev. Lett.}{\bf \ 75}, 4337 (1995).

\bibitem{14bis}  E. Santos, {\it Phys. Lett. A}, {\bf 212}, 10 (1996)

\bibitem{15}  F. De Martini, {\it Phys.Rev.Lett.{\bf 81}, 2842 (1998).}

\bibitem{16}  C. K. Hong, Z. Y. Ou, L. Mandel, {\it Phys. Rev. Lett. }{\bf 59%
}, 2044 (1987).

\bibitem{17}  G. Di Giuseppe, F. De Martini, D. Boschi, {\it Phys. Rev. A }%
{\bf 56}, 176 (1997).

\bibitem{18}  A. G. White, D. F. V. James, P. H. Eberhard, P. G. Kwiat, {\it %
Phys. Rev. Lett. }{\bf 83}, 3103 (1999).
\end{references}
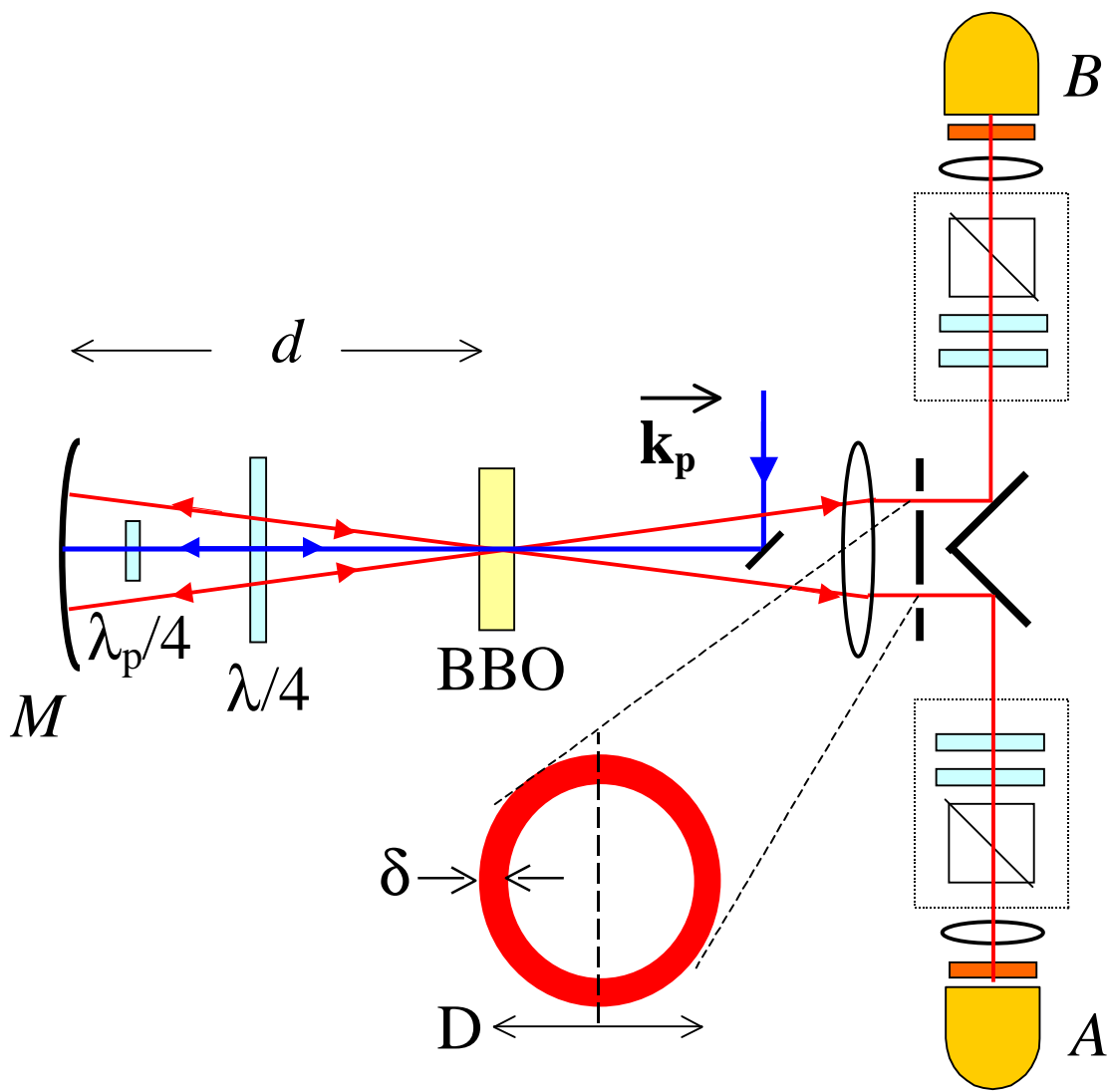
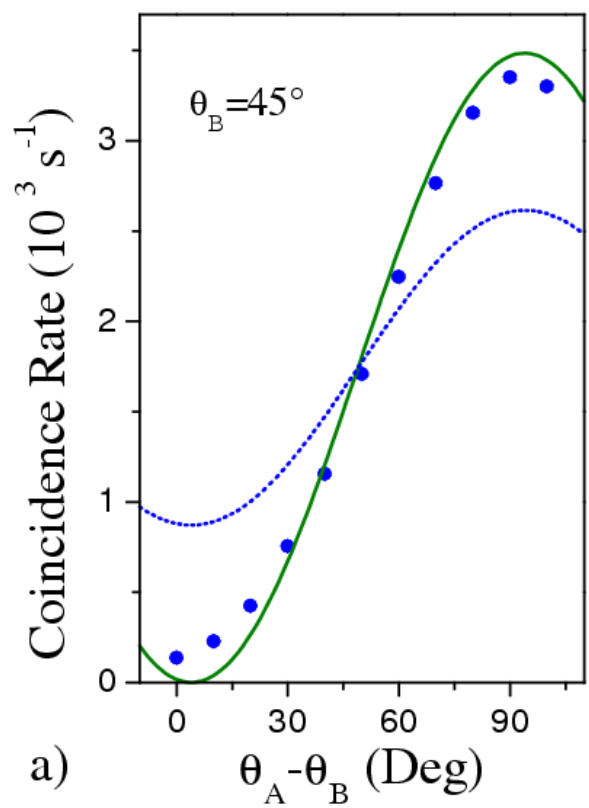
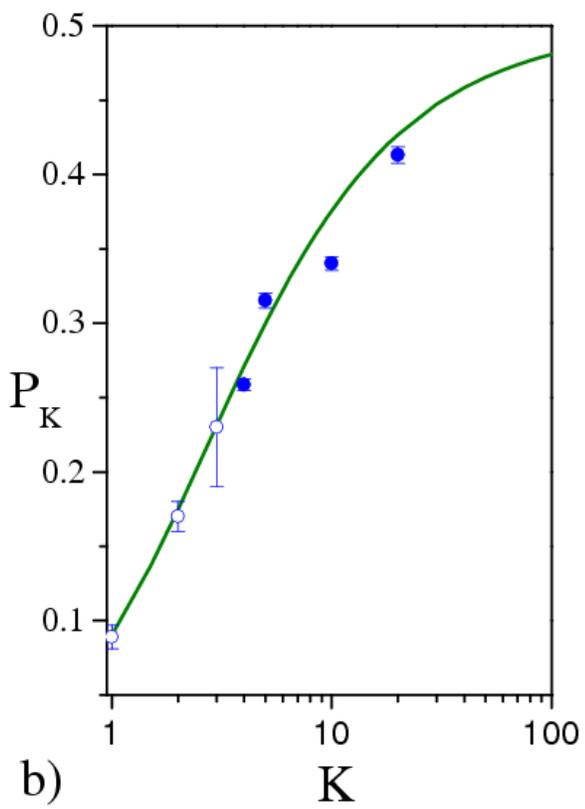
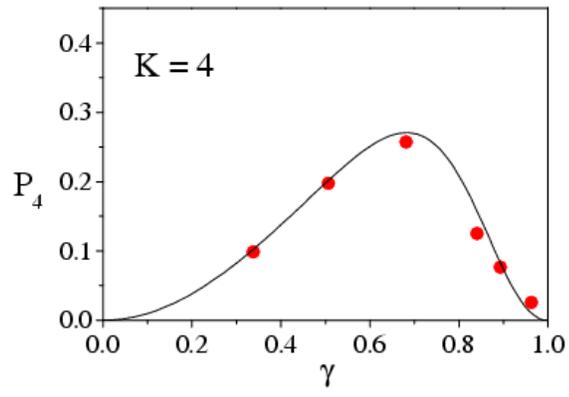
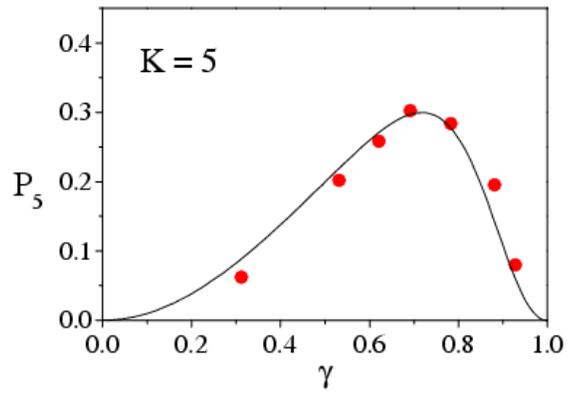
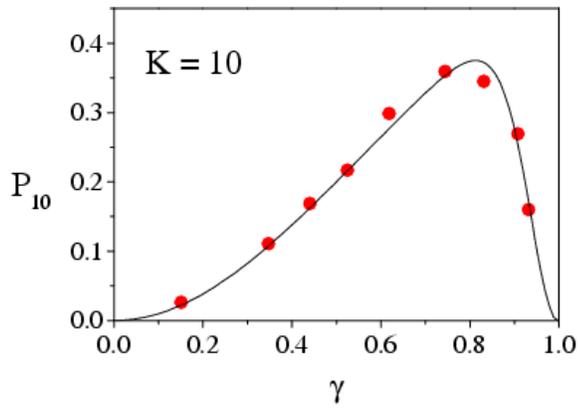
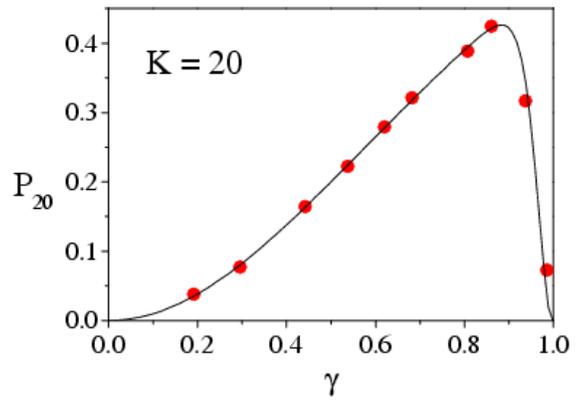
\end{document}